# Comments on "Volume ignition of mixed fuel" [1] by H. Ruhl and G. Korn (Marvel Fusion, Munich)

K. Lackner, R. Burhenn, S. Fietz, A. v. Müller

Max Planck Institute for Plasma Physics, 85748 Garching, Germany

Abstract

In the most recent note on Marvel Fusion's concept for a laser driven pB reactor without compression, Ruhl and Korn consider the volumetric energy balance of fusion reactions vs. bremsstrahlung losses in a mixed fuel (DT and pB) environment and claim the satisfaction of this necessary "ideal ignition" condition. Their results are based, however, on improper assumptions about the deposition of fusion energy in the plasma. Correcting for them, we show that the quoted composition of their fuel (a solid boron composite, binding high concentrations of D, T and p) would actually preclude ignition due to the high bremsstrahlung losses associated with the presence of boron.

To facilitate ignition, Ruhl and Korn also consider the reduction of the bremsstrahlung losses by confining the radiation in the optically thin fuel region by high Z walls. They suggest to preload this region with radiation so that the radiation temperature equals approximately that of the plasma constituents $T_r \approx T_e \approx T_i$. We show that in this set-up the radiation energy - neglected in these considerations - would, however, vastly exceed the thermal energy of the plasma and actually dominate the ignition energy requirements.

1. Introduction

In two previous notes uploaded to arXive [2,3] Ruhl and Korn have described some basic features of Marvel Fusion's concept for a laser-driven thermonuclear fusion reactor which foregoes pre-compression of the fuel and intends to utilize the aneutronic proton-boron reaction. Central to its implementation is a laser-plasma coupling scheme exploiting nano-structured targets and ultra-high-power, ultra-short pulse laser systems.

In two respective comments to their notes [4,5] we have pointed to the universality of the so-called $\varrho R$ constraint for ignition in inertial fusion systems (with $\varrho$ the density and $R$ a characteristic radius of the fuel assembly) and the resulting need for strong fuel compression, showing how it would arise also under the very different scenarios suggested by Ruhl and Korn. We also clarified that self-generated magnetic fields - at some intermediate stage suggested by the authors [2] - could not contribute to fuel confinement, due to the consequences of the virial theorem.

In a first version of a third arXive-note [6] Ruhl and Korn have addressed the hydrodynamic effects causing the disintegration of the fuel assembly, which give rise to this $\varrho R$ criterion and are generally considered the most serious constraint for inertial fusion concepts. They claimed the feasibility of "Q=1-ignition" at densities achievable in boron-hydrogen composites, albeit with "ultra-fast" laser energies in the MJ range, and relying effectively only on the DT reactions in a mixed fuel assembly. In a comment to this note [7] we pointed to inconsistencies in the assumptions leading to these estimates, showing that actual laser energies would have to be several orders of magnitude higher to initiate self-heating and sustained thermonuclear burn.

In a later version of the same arXive note [1] - subject of the present comment - the authors omitted the section on hydrodynamic effects, and restricted the consideration to pure volumetric effects: fusion plasma heating and bremsstrahlung losses.

## 2. Volume ignition in an "open" reactor design

A positive energy balance of volumetric gain and loss effects is one necessary - though in general not the most limiting - condition for ignition, generally discussed under the label "ideal ignition" [8]. For magnetic confinement systems, where particle and energy losses can to some degree be prevented by externally applied magnetic fields this has been discussed also in extended form for driven systems, non-thermal distribution functions and non-DT fusion reactions (e.g. [9], [10]).

In inertial confinement hydrodynamic effects and electron heat conduction generally pose much more stringent restrictions: in fact, for DT, except close to the minimum burn temperature, bremsstrahlung losses play only a minor role as constraint for ignition.

In the main chapter 4 of [1] Ruhl and Korn consider the question of ignition within the frame of this "ideal" model, with a positive conclusion for the specific approach taken by Marvel Fusion. As outlined in refs. [2, 3] the latter is based on the interaction of ultra-fast, ultra-powerful lasers with a nano-structured target, consisting of a boron-hydrogen composite that chemically binds high concentrations of protons, deuterons and tritons in solid, but uncompressed state. The ultimate goal is aneutronic p-$^{11}$B fusion, but for initiating burn "mixed fuels" relying on both DT and p-$^{11}$B reactions are considered. The presence of a boron matrix is an intrinsic feature of this scheme, necessary also for the initial, low-temperature DT stage, due to the envisaged scheme of laser-plasma coupling. It brings also additional, collateral benefits - like dispensing of cryogenics, and is assumed by the authors in their reference data set (used for their Fig. 5) to allow a DT mass density five times larger than that of solid, frozen DT.

We show that these positive conclusions derive exclusively from the ad-hoc assumptions for the deposition of reaction products made by Ruhl and Korn, which are in contradiction to established results of rigorous, first principle-based models. In fact, even the "ideal ignition" criterion cannot be satisfied by the Ruhl and Korn set-up due to the enhanced bremsstrahlung losses caused by the presence of boron atoms and the additional free electrons. Critical, invalid assumptions of the model of Ruhl and Korn are (1) the deposition of the full fusion energy in the plasma and (2) the description of the ion distribution by a single temperature, common both to fuels and the charged fusion products (the $\alpha$-particles).

In the low temperature range relevant for ignition and actually considered in their calculations (Fig. 5) only the DT reaction plays a role[i], in which only 20% of the fusion energy appears in charged particles that can interact with the target plasma[ii].

---

[i] For the parameters and the temperature range of Fig.5, the contribution of p-$^{11}$B reactions to the total fusion energy production is less than $10^{-3}$.
[ii] The fact, that in DT reactions 80% of the energy appears in the form of neutrons is not mentioned in the different uploaded versions [1, 6]. It is, however, of decisive importance for the ignition phase of mixed fuels.

The argument given by Ruhl and Korn for the use of a common temperature for fusion products and fuel ions is the shortness of the ion-ion equilibration time compared to other relevant time-scales. This, however, does not hold for the high energies of the $\alpha$-particles! In fact, it follows from the extensively verified theory of fast-ion slowing down that in the temperature range considered here they will deposit the dominating fraction of their energy into the electrons! – For $T_e = 9\ keV$ only a couple of percent will be deposited directly in the fuel ions (see, e.g., equ. 5.4.12 in Wesson [11]).

Correcting for these two facts completely changes the energy balance in the volume ignition model described by equ. (31) of the authors. This "ideal" model contains only fusion power as energy production and bremsstrahlung as loss mechanism. For a given composition of the plasma[iii], the former depends only on the ion, the latter only on the electron temperature. Equality of the fusion energy deposited in the plasma $P_{f,p}$ and the bremsstrahlung radiation losses $P_r$ therefore gives a relation $T_e(T_i)$ (red solid line in our Fig.1): with the region below the curve corresponding to excess fusion energy.

However, $T_e$ is not a free parameter, but will itself - for an assumed $T_i$ - be determined by the separate electron energy balance, which we can write in the form
$$P_{i,e} + f_{\alpha,e} \cdot P_{f,p} = P_b$$
where $P_{i,e}$ is the energy transferred by collisions between the fuel ions to the electrons, and $f_{\alpha,e}$ the fraction of fusion $\alpha$-particle energy transferred immediately to the electrons. This electron energy balance looks qualitatively different for Ruhl and Korn's ad-hoc model ($f_{\alpha,e} \approx 0$) and for the results of the rigorous slowing down theory ($f_{\alpha,e} \approx 1$).

Ruhl and Korn's model implies, at ignition, a collisional energy flow from ions to electrons, and implies therefore that the electrons remain cooler, which reduces the bremsstrahlung losses, and facilitates ignition (in fact, even more optimistically, they compute the bremsstrahlung losses in their Fig.5 assuming $T_e = 0.5 \cdot T_i$). For the realistic model of energy deposition on the other hand, a stationary electron energy balance will be attained, when $\alpha$-particle power matches bremsstrahlung losses, leaving no energy for transfer from electrons to bulk ions, hence corresponding to $T_e = T_i$.

---

[iii] All volumetric energy production and transfer processes considered are proportional to the square of the plasma density, and can hence be normalized, e.g. by $n_e^2$ (or by $\varrho_p^2$, as done by Ruhl and Korn).

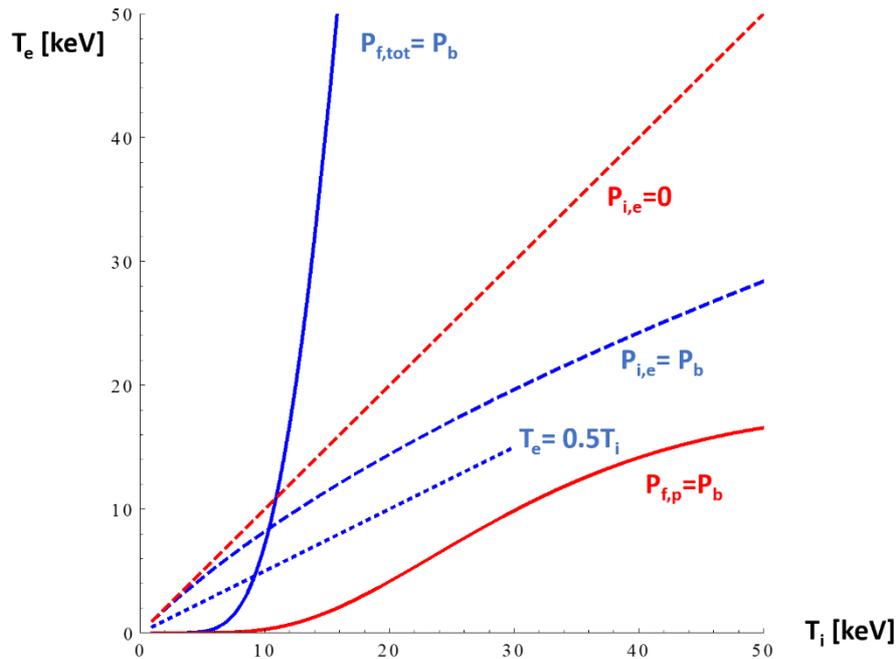

*Figure 1:* Total (solid lines) and electron (dashed lines) power balance relations determining "ideal" ignition according to the assumptions of Ruhl and Korn (blue) and to first-principle based models (red), respectively. Fusion reaction rates are taken from Bosch and Hale [12], bremsstrahlung losses from Wesson [11]. $P_{f,tot}$ is defined as total fusion power including neutrons, $P_{f,p}$ is fusion power in charged particles, $P_{i,e}$ is the power transferred from thermal ions to electrons by collisions, $P_b$ is the bremsstrahlung radiation loss. The assumptions of Ruhl and Korn allow for an intersection of both relations and hence ignition: the first principle based model does not. - In the actual calculations shown in Fig. 5 of ref. [1] the even more optimistic assumption $T_e=0.5T_i$ (dotted line) is used as substitute for the electron power balance.

The relations determining marginal ignition conditions are summarized in the $T_e$ vs. $T_i$ - plane in Fig.1., with the ab-initio based results (only charged fusion products contributing to plasma heating and $\alpha$-particle energy nearly exclusively deposited into electrons) shown by the red, and the assumptions of Ruhl and Korn (also neutron energy deposited in plasma, total fusion energy into ions) by the blue curves. Whereas the curves corresponding to the assumptions of Ruhl and Korn cross in the 10 keV region, and hence allow „volume ignition", in reality bremsstrahlung losses will always exceed fusion plasma heating for the plasma composition of their target set-up. "Ideal" or volume ignition, defined by a crossing of the solid and dashed curves would be readily attainable under the assumptions of Ruhl and Korn (blue curves), but is impossible based on the first principle-based results.

The impossibility of attaining even "ideal ignition" in a scenario relying essentially on DT for fusion power production might appear somewhat surprising: it is due to the enhanced bremsstrahlung losses linked to the presence of boron, which in turn is motivated by their particular laser deposition scheme. For $T_e = T_i$, the equimolar mix used in the ignition estimates of Ref. [1] (e.g., Fig.5)[iv] implies an enhancement of bremsstrahlung viz. fusion power by a factor of 56 compared to a pure DT fuel! The five-fold increase in DT density in their set-up compared to conventional DT ice has no effect on volume ignition, as for the latter all competing effects depend in equal form on density (only their relative composition has an effect).

---

[iv] The composition of the boron compound described by Ruhl and Korn corresponds to diborane ($B_2H_6$), which is, however, a gas. Boron based solids with a large capacity for hydrogen storage do exist, but involve also other elements, like borazane ($H_3NBH_3$). In this case, however, the presence of N would raise the bremsstrahlung losses by a further 60% and reduce the p-$^{11}$B yield by a factor of 2.

## 3. A "closed" reactor design

Even within the frame of their ad-hoc assumptions, the "ideal ignition" model of Ruhl and Korn would result in very high ignition energies in the 10s of Megajoule range (equ. 34) which would have to be provided for by an extremely advance short pulse laser system. They therefore consider means to reduce the bremsstrahlung radiation losses - which according to the first-principle based model would actually be absolutely mandatory. They propose therefore a "closed reactor" configuration, in which the bremsstrahlung radiation would be largely confined by reflection from high-Z walls, immersing the target in a radiation field with a radiation temperature $T_r$. Once equality between the electron and the radiation temperature were established, absorption would compensate for emission, even if the plasma were optically thin.

This concept implies enclosing the fusion plasma in a hohlraum cavity, with the radiation temperature of the latter equal to $T_e$. Ignition would thus require not only to heat the plasma, but also the photons in the cavity to $T_r = T_e$, "preloading it with radiation". This hohlraum concept is actually well known and tested in indirect drive fusion, albeit at radiation (equal to wall) temperatures in the 300 eV, rather than in the multi-keV range. In fact, at the latter (fusion) temperatures we would enter a totally new regime, as the usual assumption of neglecting the photon energy in the cavity energy balance (chapter 9.3.1. of ref. [8]) would cease to be valid! For the electron density of the reference case of ref. [1] (Fig. 5), the photon energy would equal the thermal electron kinetic energy for $T_r = T_e \approx 3\ keV$, and due to its rise $\sim T^4$ exceed it for the reference electron temperature of 8.5 keV by a factor of 22 (!). This would hold for a completely tight-fitting cavity shell, and no voids in the fuel assembly (contrary to the sketches in Figs. 1 and 6), which otherwise would also have to be filled with radiation energy. The contribution of the energy in the radiation field to the ignition energy, which would actually dominate, is not considered by the authors.

## 4. A dynamic model of thermonuclear burn

In addition to these marginal equilibrium considerations, Ruhl and Korn also describe a dynamic model of thermonuclear burn (equations (35)). It is zero-dimensional, and describes the time-dependence in terms of a finite difference formulation. In particular, however, it repeats their ad-hoc assumptions concerning fusion power deposition of the equilibrium model. Fuel ions and fusion-products are assumed to have the same ion temperature and the total fusion energy (including the 80% in the DT produced neutrons) is instantaneously transferred to them. Energy is transferred from the fuel ions by collision to electrons and subsequently lost by radiation. This model, of course, dramatically overestimates plasma heating (in reality the neutrons will heat the walls - or, in an extremely large fuel assembly the distant, cold layers) and underestimates the bremsstrahlung losses. It hence can also predict ignition, in a situation where this is already ruled out by the first-principle based model for volume ignition.